\documentstyle[a4,11pt]{article}
\begin{document}

\setlength{\baselineskip}{26pt}

\noindent
{\Large \bf Earth Similarity Index with two free parameters}

\vspace{4mm}

\setlength{\baselineskip}{18pt}

\noindent
{\large Suresh Chandra$^1$, Subas Nepal$^1$ and Mohit K. Sharma$^2$}

\noindent
$^1$Physics Department, Lovely Professional University, Phagwara 144411
(Punjab), India

\noindent
$^2$School of Studies in Physics, Jiwaji University, Gwalior 474011 (M.P.),
India

\noindent
Emails: suresh492000@yahoo.co.in; nsubas11@gmail.com; mohitkumarsharma32@yahoo.in

\vspace{6mm}

\noindent
{\bf Abstract.} We have derived Earth Similarity Index (ESI) with two free parameters $m$
 and $T$. These free parameters are optimized with the consideration that the planet Mars is 
almost similar to the Earth. For the optimized values of free parameters, the 
interior-ESI, surface-ESI and ESI for some planets are calculated. The results for
 $m = 0.8$ and $T = 0.8$  are compared with the values obtained by Schulze-Makuch 
{\it et al.} (2011). We have found that the exoplanet 55 Cnc f is within 10\% away from 
the threshold value $T$. The exoplanets HD 69830 c, 55 Cnc c, 55 Cnc f, 61 Vir d and HIP
 57050 b are found to have ESI within 10\% from the threshold value.  


\noindent

\section{Introduction}

In 1584, Catholic Monk Giordano Bruno asserted that there are countless suns and 
countless earths, and all these earths are revolving around their respective suns. When
he said this in public, he was removed from the Church by charging that he was against 
the religion. But, this fact was found true when the confirmation of the first extra-solar
planet was announced by Mayor \& Queloz (1995), revolving around a sun-type. 51 Pegasi. 
Detection of new planets is progressing continuously and as of 9 July 2015, there are
 1858 exoplanets, including 468 multiple planetary
systems (http://exoplanetarchive.ipac.caltech.edu/).

The number of known exoplanets is increasing each day as the detection methods used
for both the ground based and space missions have improved in terms of technology and 
more scientists are getting interested in this field. There are many ground based and 
space missions to detect the exoplanets. The Search for Extra Terrestrial Intelligence 
(SETI) is the name of mission for searching life-forms outside the Earth. Radio 
telescopes having large antennas are being used to investigate the exoplanets. In 
Arizona, the Large Binocular Telescope Interferometer (LBTI) is being used for advanced 
interferometry for detection of exoplanets. These instruments are bound to advance and 
improve our understanding and detection of a number of exoplanets which may be similar to
 our Earth or not. The Kepler mission is aimed for detection of earth-like planets and 
the space missions such as CHEOPS (CHaracterizing ExOPlanet Satellite) and James Webb 
Space Telescope (JWST) are in progress. With the new generation of telescopes and 
missions, detection of a large number of exoplanets is expected. The exoplanets need to 
be classified if one is similar to the Earth or not. We therefore need a scheme which can
 be used for classification on the basis of data available to us. The data contain the 
information about planets, such as mass, radius, surface temperature, etc. These 
planetary parameters can be used to know about the probability of existing life-forms 
some where other than the Earth. The probability of existing life-forms outside the Earth
 is given as a concept of 'circumstellar habitable zone' by Kasting {\it et al.} (1993). 
It suggests that the focus should be made on those worlds which can hold atmosphere and 
liquid water. Some life-forms are most likely to be found on those planetary bodies which
 have similar Earth-like conditions.

A two-tiered classification scheme of exoplanets is suggested by
Schulze-Makuch {\it et al.} (2011). The first tier consists of Earth Similarity Index
(ESI), which decides about the worlds with respect to their similarity to the Earth. The
ESI was calculated on the basis of data for exoplanets, such as mass, radius and
temperature. The second tier of the scheme is the Planetary Habitability Index
(PHI), based on the presence of a stable substrate, available energy, appropriate
chemistry and potential for holding liquid solvent.

In the present investigation, we have derived the ESI with two free parameters. These 
free parameters have been optimized with the consideration that the planet Mars is almost
 similar to the Earth. For the optimized values of free parameters, the interior-ESI, 
surface-ESI and ESI for some planets are calculated. The results are compared with the 
available data.

\section{Earth similarity index}

The similarity index is a mathematical tool that can be applied for a set of data. 
It is used in various fields of science, such as Mathematics ({\it e.g.}, Set
Theory), Ecology ({\it e.g.}, Sorensen similarity index), Computer imaging ({\it e.g.},
structural similarity index) and many others (Schulze-Makuch {\it et al.}, 2011). This 
method is a measure of deviation from a reference system, usually on a scale lying 
between zero and one. The ESI is a quantitative measure of Earth-likeness.

The ESI, we have proposed, may be understood in the following manner. Consider a 
physical quantity having value $x_0$ on the surface of our Earth. Suppose, the value of
this quantity varies from $x_a$ to $x_b$, such that $x_a < x_0 < x_b$, on the Earth. 
Then, the percentage variations $p$ and $q$ are expressed as
\begin{eqnarray}
p = \frac{x_0 - x_a}{x_0} \ \times 100 \hspace{1.5cm} \mbox{and}  \hspace{1.5cm}
q = \frac{x_b - x_0}{x_0}  \ \times 100 \nonumber    
\end{eqnarray}

Now, we define a threshold value $T$ as
\begin{eqnarray}
T = \Big[1 - \Big(\frac{x_0 - x_a}{x_0 + x_a}\Big)^m\Big]^{w_a} \label{eq03}
\end{eqnarray}

\noindent
and
\vspace{-3mm}
\begin{eqnarray}
T = \Big[1 - \Big(\frac{x_b - x_0}{x_b + x_0}\Big)^m\Big]^{w_b} \label{eq04}
\end{eqnarray}

\noindent
where $w_a$ and $w_b$ denote the weight exponents and $m$ is a free parameter. The 
threshold is the limiting value of ESI above which a planet is considered similar
to the Earth. Further, the threshold value $T$ can be expressed as
\begin{eqnarray}
T = \Big(1 - \frac{t}{100}\Big) \nonumber 
\end{eqnarray}

\noindent
where $t$ can assume positive values between zero and 100, so that $T$ is positive, 
having the value between zero and 1. Thus, we have taken $T$ also as another free 
parameter. In the work of Schulze-Makuch {\it et al.} (2011), $t$ is taken as 20, so that
 $T$ = 0.8, and $m$ as 1. In our investigation, we have considered $m$ to assume the 
values from 0.6 to 1.2. 

From equation (\ref{eq03}) and (\ref{eq04}), we get
\begin{eqnarray}
w_a = \frac{\mbox{ln} T}{\mbox{ln} [ 1 - (\frac{p}{200-p})^m]}  \hspace{1.5cm} \mbox{and}
  \hspace{1.5cm} w_b = \frac{\mbox{ln} T}{\mbox{ln} [ 1 - (\frac{q}{200+q})^m]} \nonumber
\end{eqnarray}

\noindent
We finally take the weight exponent $w_x$ as the geometrical mean of $w_a$ and $w_b$.
That is
\begin{eqnarray}
 w_x = \sqrt{w_a \times w_b}  \nonumber    
\end{eqnarray}

\noindent
Here, the subscript $x$ denotes a physical quantity.

The basic $ESI_x$ for a physical quantity $x$ is expressed as 
\begin{eqnarray}
ESI_x = \Big(1 - \Big|\frac{x-x_0}{x+x_0}\Big|^m\Big)^{w_x} \nonumber
\end{eqnarray}

\noindent
For the planetary property $x$, the terrestrial reference is $x_0$. The $ESI_x$ can 
assume a value between zero (no similarity with the Earth) and one (identical to the
Earth). We have calculates $ESI_r$, $ESI_\rho$, $ESI_e$ and $ESI_T$ corresponding to 
radius, density, escape velocity and temperature, respectively. 

The $ESI_r$ and $ESI_\rho$, for mean radius $r$ and bulk density $\rho$, respectively, 
are used to define the interior earth similarity index $ESI_I$, expressed as
\begin{eqnarray}
ESI_I = (ESI_r \times ESI_\rho)^{1/2} \label{eq10}
\end{eqnarray}

\noindent
Thus, the $ESI_I$ is the geometrical mean of $ESI_r$ and $ESI_\rho$. The  $ESI_e$ and
$ESI_T$ for escape velocity $v$ and mean surface temperature $T$, respectively, are
used to define the surface earth similarity index $ESI_S$, expressed as
\begin{eqnarray}
ESI_S = (ESI_e \times ESI_T)^{1/2} \label{eq11}
\end{eqnarray}

\noindent
Thus, the $ESI_S$ is the geometrical mean of $ESI_e$ and $ESI_T$. The ESI (sometimes 
called the global ESI) is expressed as
\begin{eqnarray}
ESI = (ESI_I \times ESI_S)^{1/2} \label{eq12}
\end{eqnarray}

\noindent
Using equations (\ref{eq10}) and (\ref{eq11}) in (\ref{eq12}), we have
\begin{eqnarray}
ESI = (ESI_r \times ESI_\rho \times ESI_e \times ESI_T)^{1/4} \nonumber
\end{eqnarray}
 
\section{Analysis}

In our investigation, we have considered two values of $T$ as 0.8 and 0.9. The value of 
$T$ can never be greater than 1, as we are considering the Earth to be the most superior 
planet. The variations of parameters, relative to those of the mean values on the Earth,
 are taken as the following.  The definitional limits for radius are from 0.5 to 1.5 
times the Earth's radius (Sotin {\it et al.}, 2009). The limit for the mass of an
existing planet is from 0.1 to 10 times that of the Earth (Gaidos {\it et al.},
2005). For the density, the definitional limits  are 0.7 and 1.5 times the Earth's
density. The temperature variation is taken from 273 K to 232 K. The definitional limits
 for the escape velocity are considered as 0.4 and 1.4 times that of the Earth. For 
these limits, we have calculated earth similarity indexes corresponding to 
radius, density, escape velocity and temperature.

For the given values of parameters for the planets in our solar system, we have 
calculated  ESI, where $m$ and $T$ are considered as the free parameters. The values of 
ESI are found large for the Mars, Mercury and Venus. For these three planets, in Figure 
\ref{fig1},  we have plotted ESI versus $m$ for $T$ = 0.8 and $T$ = 0.9. In the figure, 
solid line is for the Mars, dashed line for the Mercury and dotted line for the Venus. 
For other solar planets, the  ESI is small and therefore is not shown in the figure. We 
have also drawn a  line for ESI = 0.8 and ESI = 0.9 in the respective figure. For the 
optimization of free parameter $m$, we have considered that the Mars has the ESI equal to
 the terminal value $T$. For both, $T$ = 0.8 and $T$ = 0.9, we have found $m$ = 0.8, 
where the ESI of Mars is equal to the terminal value $T$. Hence, in the further 
calculations of the $ESI_S$, $ESI_I$ and ESI for various solar planets and satellites, 
and exoplanets, we have taken two values of $T$ as 0.8 and 0.9, and the value of $m$ as 
0.8. 

In Table 1, we have given the values of physical parameters, interior-ESI, surface-ESI
and ESI for 31 objects. For 21 objects, the values of physical parameters are the
same as given by Schulze-Makuch {\it et al.} (2011). In their calculations, there are
$m$ = 1 and $T$ = 0.8. In the last column of Table 1, we have given the values of
Schulze-Makuch {\it et al.} (2011) for the ESI.
Table 2 gives the similar results for $m$ = 0.8 and $T$ = 0.9. 

Most of the exoplanets are detected by transit photometry method and radial velocity
method. The transit photometry method cannot measure the mass of exoplanet whereas the
radial velocity method cannot measure the radius of exoplanet (Jones {\it et al.}, 2008).
 So, for the exoplanets for which either mass or radius was not measured, the mass or 
radius is calculated with the help of mass-radius relation, given by Sotin {\it et al.} 
(2007).

\section{Discussion}

From Table 1, we have found that for each object, the present value of ESI, in
general, larger than that of Schulze-Makuch {\it et al.} (2011). It is due to the
change in the value of $m$.  We have also found that the exoplanet 55 Cnc f is within 
10\% limit from the threshold value $T$. The exoplanets HD 69830 d, 55 Cnc c, 55 Cnc f, 
61 Vir d and HIP  57050 b have the ESI within 10\% limit from the threshold value.
This supports the opinion for existence of life in the universe.

\section{Conclusion}

We have derived the ESI with two free parameters, the threshold value $T$ and $m$. After
 optimization that the ESI of Mars is equal to the threshold value $T$, we got the value 
of $m$ as 0.8. In the paper of Schulze-Makuch {\it et al.} (2011), the value of $m$ is 1.
 For the present value of $m$ = 0.8, some exoplanets have come in the range up to the 
threshold value. It enhances the probability of finding life in the interstellar medium. 

\vspace{5mm}

\newpage

\noindent
{\bf Acknowledgments}  

Financial support from the Department of Science \& Technology is thankfully 
acknowledged. Thanks are due to Lovely Professional University, Punjab for 
encouragements.

\vspace{6mm}

\noindent
{\large\bf  References}

\begin{description}

\item{} Gaidos, E., Deschenes, B., Dundon, L. and Fagan, K. (2005) Beyond the principle
of plentitude: a review of terrestrial planet habitability. Astrobiology 5: 100-126.

\item{} Jones, B. (2008) Exoplanets search methods, discoveries, and prospects for 
astrobiology. International Journal of Astrobiology 7: 279-292. 

\item{} Kasting, J.F. (1993) Earth's early atmosphere. Science 259: 920 - 926.

\item{} Mayor, M. and Queloz, D. (1995) A Jupiter-mass companion to a solar type star.
Nature 378: 355 - 359.

\item{} Schulze-Makuch, D., Mendez, A., Fairen, A.G., von Paris, P., Turse, C., Boyer, 
G., Davila, A.F., Antonio, M.R.D.S., Catling, D. and Irwin, L.N.  (2011)
A two-tiered approach to assess the habitability of exoplanets. Astrobiology 11:
1041 - 1052.

\item{} Sotin, C., Grasset, O. and Mocquet, D. (2007) Mass-radius curve for extrasolar 
earth like planets and ocean planets, Icarus 191: 337 - 351.  

\end{description}

\vspace{4mm}

\hspace{-17mm}
\begin{tabular}{llrccrcccc}
\multicolumn{10}{l}{Table 1. Values of various ESI for some solar planets and satellites
and exoplanets  with $m$ = 0.8} \\
\multicolumn{10}{l}{and $T$ = 0.8.} \\
\hline
\multicolumn{1}{c}{S.No.} & \multicolumn{1}{c}{Body} & \multicolumn{1}{c}{Radius} 
& \multicolumn{1}{c}{Density} & \multicolumn{1}{c}{Esc. Vel.} & \multicolumn{1}{c}{Temp} 
& \multicolumn{1}{c}{$ESI_I$} & \multicolumn{1}{c}{$ESI_S$} & \multicolumn{1}{c}
{ESI} & \multicolumn{1}{c}{ESI*} \\
 & & \multicolumn{1}{c}{(EU)} & \multicolumn{1}{c}{(EU)} & \multicolumn{1}{c}{(EU)} & 
\multicolumn{1}{c}{K} & &  & &  \\
\hline
   1 &   Earth        &    1.000 &    1.000 &    1.000 &     288. &   1.0000 &   1.0000 &   1.0000 &  1.000 \\
   2 &   Mars         &    0.532 &    0.713 &    0.449 &     227. &   0.8643 &   0.7394 &   0.7994 &  0.700 \\
   3 &   Mercury      &    0.382 &    0.984 &    0.379 &     440. &   0.8773 &   0.6143 &   0.7341 &  0.600 \\
   4 &   Moon         &    0.272 &    0.606 &    0.212 &     220. &   0.7509 &   0.6235 &   0.6842 &  0.560 \\
   5 &   Venus        &    0.949 &    0.950 &    0.925 &     730. &   0.9849 &   0.4381 &   0.6568 &  0.440 \\
   6 &   Io           &    0.285 &    0.639 &    0.228 &     130. &   0.7660 &   0.3941 &   0.5494 &  0.360 \\
   7 &   Callisto    &    0.378 &    0.352 &    0.218 &     134. &   0.6916 &   0.4022 &   0.5274 &  0.340 \\
   8 &   Jupiter      &   10.973 &    0.240 &    5.379 &     152. &   0.4800 &   0.4387 &   0.4589 &  0.290 \\
   9 &   Ganymede     &    0.412 &    0.352 &    0.244 &     110. &   0.7009 &   0.3375 &   0.4864 &  0.290 \\
  10 &   Ceres        &    0.074 &    0.376 &    0.045 &     167. &   0.5225 &   0.3447 &   0.4244 &  0.270 \\
  11 &   Europa       &    0.244 &    0.546 &    0.180 &     102. &   0.7194 &   0.2925 &   0.4587 &  0.260 \\
  12 &   Saturn       &    9.140 &    0.124 &    3.225 &     134. &   0.4049 &   0.4318 &   0.4181 &  0.250 \\
  13 &   Titan        &    0.404 &    0.341 &    0.235 &      94. &   0.6928 &   0.2836 &   0.4432 &  0.240 \\
  14 &   Uranus       &    3.980 &    0.230 &    1.910 &      76. &   0.5728 &   0.2604 &   0.3862 &  0.190 \\
  15 &   Neptune      &    3.864 &    0.297 &    2.105 &      72. &   0.6200 &   0.2409 &   0.3865 &  0.180 \\
  16 &   Titania      &    0.123 &    0.301 &    0.069 &      60. &   0.5433 &   0.1295 &   0.2652 &  0.100 \\
  17 &   Enceladus    &    0.039 &    0.291 &    0.021 &      75. &   0.4271 &   0.1255 &   0.2315 &  0.094 \\
  18 &   Pluto        &    0.180 &    0.371 &    0.109 &      40. &   0.6180 &   0.0878 &   0.2329 &  0.075 \\
  19 &   Triton       &    0.212 &    0.379 &    0.130 &      38. &   0.6402 &   0.0856 &   0.2341 &  0.074 \\
  20 &   HD69830 d    &    4.190 &    0.250 &    2.100 &     312. &   0.5818 &   0.8439 &   0.7007 &  0.600 \\
  21 &   55 Cnc c     &    5.680 &    0.250 &    2.840 &     310. &   0.5510 &   0.8031 &   0.6652 &  0.560 \\
  22 &   55 Cnc f     &    4.910 &    0.390 &    3.060 &     310. &   0.6405 &   0.7918 &   0.7121 &  0.614 \\
  23 &   61 Vir d     &    3.680 &    0.460 &    2.500 &     375. &   0.7028 &   0.7093 &   0.7060 &  \\
  24 &   HIP 57050 b  &    6.640 &    0.320 &    3.780 &     250. &   0.5743 &   0.7217 &   0.6438 &  \\
  25 &   Mu Ara d     &    8.380 &    0.280 &    4.450 &     327. &   0.5292 &   0.7059 &   0.6112 &  \\
  26 &   HD 142 b     &   11.120 &    0.240 &    5.430 &     286. &   0.4787 &   0.7397 &   0.5951 &  \\
  27 &   HD 96167 b   &    9.360 &    0.260 &    4.810 &     334. &   0.5070 &   0.6833 &   0.5886 &  \\
  28 &   HD 108874 b  &   12.490 &    0.220 &    5.890 &     294. &   0.4558 &   0.7197 &   0.5728 &  \\
  29 &   HD 210277 b  &   11.980 &    0.230 &    5.720 &     275. &   0.4658 &   0.7109 &   0.5754 &  \\
  30 &   HD 147513 b  &   10.990 &    0.240 &    5.380 &     263. &   0.4798 &   0.6970 &   0.5783 &  \\
  31 &   HD 69830 c   &    2.820 &    0.540 &    2.070 &     549. &   0.7636 &   0.5253 &   0.6333 &  \\
\hline
\multicolumn{10}{l}{*Values reported by Schulze-Makuch
{\it et al.} (2011) for $m$ = 1 and $T$ = 0.8.} \\
\end{tabular}

\begin{tabular}{llccc}
\multicolumn{5}{l}{Table 2. Same as Table 1 with $m$ = 0.8 and $T$ = 0.9.} \\
\hline
\multicolumn{1}{c}{S.No.} & \multicolumn{1}{c}{Body} & \multicolumn{1}{c}{$ESI_I$} & 
\multicolumn{1}{c}{$ESI_S$} & \multicolumn{1}{c} {ESI} \\
\hline
   1 &   Earth        &   1.0000 &   1.0000 &   1.0000 \\
   2 &   Mars         &   0.9334 &   0.8671 &   0.8997 \\
   3 &   Mercury      &   0.9401 &   0.7945 &   0.8642 \\
   4 &   Moon         &   0.8735 &   0.8001 &   0.8360 \\
   5 &   Venus        &   0.9928 &   0.6773 &   0.8200 \\
   6 &   Io           &   0.8817 &   0.6442 &   0.7537 \\
   7 &   Callisto    &   0.8402 &   0.6505 &   0.7393 \\
   8 &   Jupiter      &   0.7071 &   0.6777 &   0.6923 \\
   9 &   Ganymede     &   0.8455 &   0.5988 &   0.7115 \\
  10 &   Ceres        &   0.7360 &   0.6048 &   0.6672 \\
  11 &   Europa       &   0.8560 &   0.5597 &   0.6922 \\
  12 &   Saturn       &   0.6525 &   0.6726 &   0.6625 \\
  13 &   Titan        &   0.8409 &   0.5515 &   0.6810 \\
  14 &   Uranus       &   0.7686 &   0.5298 &   0.6381 \\
  15 &   Neptune      &   0.7980 &   0.5106 &   0.6383 \\
  16 &   Titania      &   0.7497 &   0.3809 &   0.5344 \\
  17 &   Enceladus    &   0.6692 &   0.3753 &   0.5012 \\
  18 &   Pluto        &   0.7968 &   0.3170 &   0.5026 \\
  19 &   Triton       &   0.8101 &   0.3133 &   0.5038 \\
  20 &   HD69830 d    &   0.7744 &   0.9230 &   0.8454 \\
  21 &   55 Cnc c     &   0.7547 &   0.9016 &   0.8249 \\
  22 &   55 Cnc f     &   0.8103 &   0.8956 &   0.8519 \\
  23 &   61 Vir d     &   0.8466 &   0.8503 &   0.8484 \\
  24 &   HIP 57050 b  &   0.7696 &   0.8573 &   0.8123 \\
  25 &   Mu Ara d     &   0.7404 &   0.8483 &   0.7926 \\
  26 &   HD 142 b     &   0.7062 &   0.8673 &   0.7826 \\
  27 &   HD 96167 b   &   0.7256 &   0.8354 &   0.7786 \\
  28 &   HD 108874 b  &   0.6901 &   0.8562 &   0.7686 \\
  29 &   HD 210277 b  &   0.6971 &   0.8512 &   0.7703 \\
  30 &   HD 147513 b  &   0.7070 &   0.8433 &   0.7721 \\
  31 &   HD 69830 c   &   0.8804 &   0.7379 &   0.8060 \\
\hline
\end{tabular}

\begin{figure}[h]
\vspace{0mm}
\vspace{10mm}
\caption{\small Variation of ESI versus $m$ for $T$ = 0.8 and $T$ = 0.9. Solid line is 
for the Mars, dashed line for the Mercury and dotted line for the Venus. We have also 
plotted line for ESI = 0.8 and ESI = 0.9 in the respective figure. For other 
solar-planets, the value of ESI is very small.} \label{fig1}
\end{figure}

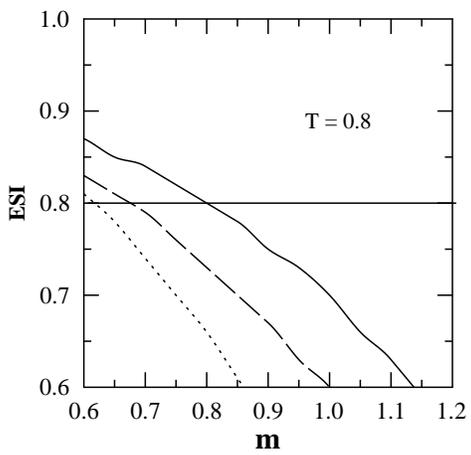
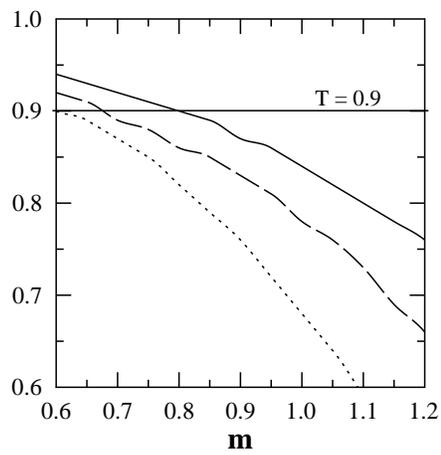
\end{document}